\documentclass[aps,prl,twocolumn,superscriptaddress]{revtex4-1}
\usepackage{graphicx}
\usepackage{dcolumn}
\usepackage{bm}

\usepackage{amsmath}
\usepackage[english]{babel}
\usepackage{graphicx}
\usepackage{dcolumn}
\usepackage{bm}
\newcommand{\deriv}{\mathrm{d}}
\usepackage{xcolor}

\begin{document}
\title{Multiple scattering enabled superdirectivity \\ from a subwavelength ensemble of resonators}

\author{S. Metais}
\affiliation{Institut Langevin, ESPCI Paris, PSL Research University, CNRS UMR 7587,  1 rue Jussieu, 75005 Paris, France}
\author{G. Lerosey}
\affiliation{Greenerwave, ESPCI Paris Incubator PC'up, 6 rue Jean Calvin, 75005 Paris, France}
\author{ F. Lemoult}
\email{fabrice.lemoult@espci.psl.eu}
\affiliation{Institut Langevin, ESPCI Paris, PSL Research University, CNRS UMR 7587,  1 rue Jussieu, 75005 Paris, France}

\date{\today}

\begin{abstract}
An ensemble of resonators arranged on a sub-wavelength scale is usually considered as a bulk effective medium, known as a metamaterial, and can offer unusual macroscopic properties. Here, we take a different approach and limit ourselves to the study of only a few number of such elementary components and demonstrate that it still offers uncommon opportunities. Typically, thanks to the multiple scattering and the phase shift that the resonances offer, we observe fields that vary at scales completely independent of the wavelength in freespace. By smartly tuning the resonance frequencies, we can design at will the complex current distribution in those resonators. This way, we design a superdirective antenna, {\it ie.} an antenna that is surprisingly more directive than its size would foreshadow. This approach is verified numerically and experimentally in the context of microwaves, but this applies to any wave-field where sub-wavelength resonators exist.
\end{abstract}

\maketitle
Focusing wave with the use of a microscope lens is subject to the Abbe's limit~\cite{abbe} which relates the focal width to the aperture. In the 1950s, \citet{di_francia_super-gain_1952} demonstrated that the focal width of an optical lens can actually be decreased below the Abbe's limit if you add a well-controlled apodization to the aperture. This effect is now known under the name of super-oscillation~\cite{huang2009super}, and focal widths thinner than the diffraction limit of $\lambda/3.5$ have been experimentally verified in optics~\cite{rogers2012super,chen_superoscillation:_2019}.

In the context of antennas, the equivalent of the notion of focusing is actually performed in the far-field region, and the equivalent of the Abbe's limit now links the size of an antenna to its achievable directivity, {\it ie.} its ability to concentrate the radiated energy solely in one direction. By decomposing the radiated field onto the natural basis of the spherical harmonics, we similarly obtain the Chu's limit~\cite{chu_physical_1948,harrington_effect_1960}, the direct analogue of the Abbe's limit. The equivalent of the super-oscillation was theoretically proposed by~\citet{schelkunoff_mathematical_nodate} when he showed, actually slightly earlier, that a linear array of antennas can overcome this limit and behaves as a so called \emph{super-directive} antenna. This theoretical prediction remains  a wide-open field. As an exemple, there have been some proposals such as the use a very high index dielectric material, in order to generate higher multipolar contributions to the farfield~\cite{krasnok_superdirective_2014} or the tuning of a metasurface from super-resolution to super-directivity~\cite{ludwig_metascreen-based_2012}. Nowadays, in the context of miniaturization, it would be interesting to build such compact directive antennas with an overall dimension comparable to the wavelength. But now, we face the problem that in the~\citet{schelkunoff_mathematical_nodate} configuration all antennas are strongly correlated due to the sub-wavelength spacing. 

It happens actually that we fall within the scope of metamaterials~\cite{bookCapolino,BookStefanMaier}, which are media that contains inclusions organized on a subwavelength scale. Notably, if we consider a linear array of half-wavelength-long metallic wires, as in the geometry of~\citet{schelkunoff_mathematical_nodate}, but arranged with a separation distance of $d$ satisfying the condition $d\ll\lambda$ ($\lambda$ being the freespace wavelength), we have a metamaterial known as the uniaxial wire medium~\cite{Belov2002}. The propagation within such a medium is well modelled by a Lorentz dispersion model~\cite{tretyakov2005}.  Typically, below the resonance frequency of the wires, the medium exhibits a high effective permittivity, and the propagating modes spatially oscillate on scales smaller than the vacuum wavelength. Those modes have been used to achieve focusing below the freespace diffraction limit from the farfield in a two-dimensional array of resonators~\cite{lerosey_focusing_2007,lemoult_resonant_2010,LemoultWRM2011b}.

\begin{figure*}[bt]
\begin{center}
 \includegraphics[width=17cm]{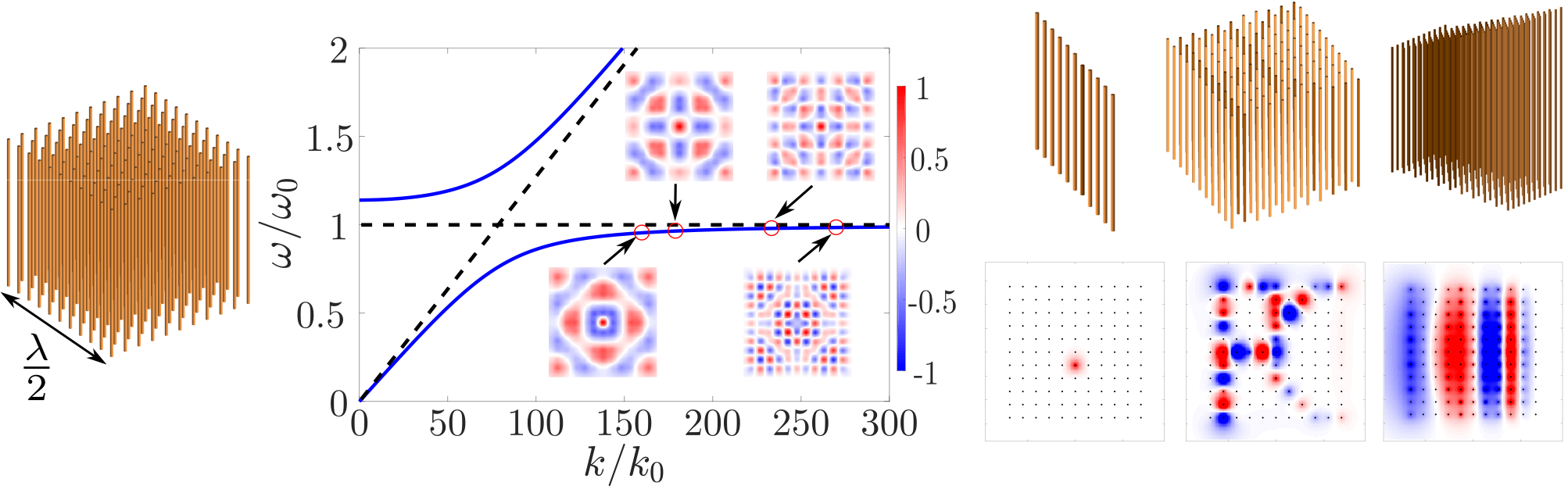}
 \caption{Simulation of a two dimensional square lattice arrangement of half-wavelength long wires. The medium is hence a cube of $\lambda/2$ side, as shown on the left. The dispersion relation of the infinite medium is a polariton one, as shown in the center, which exhibits, as the frequency approaches $f_0$, very high wave vectors. The insets show normalized near-field calculations of the plane above the medium for the frequency of the associated red circles. Defects in this homogeneous medium allows one to control the field with the defect's scale. The top part shows the medium, with respectively, one shorter wire, a line of shorter wires, and a gradient of lengths, and the bottom part shows the corresponding near fields : a cavity, a wave-guide, and a rainbow trapping effect.  } \label{fig_2D}
\end{center}
\end{figure*}

Another way to interpret those dispersion curves is to see each individual wire as a resonant scatterer~\cite{Lagendijk1993,Ad1998}. The macroscopic wave which propagates within the medium is therefore the consequence of all the multiple scattering events occurring between those resonators, albeit the subwavelength dimensions at play~\cite{kaina_negative_2015}. In this letter, we will therefore demonstrate that it is possible to tune each wire in order to turn a subwavelength collection of those wires onto a superdirective antenna. Instead of using a metamaterial for focusing below the diffraction limit as initially proposed by~\citet{pendry_negative_2000}, we ironically use it in order to beat the farfield equivalent, namely the Chu's limit~\cite{chu_physical_1948}.  After explaining our principle theoretically, we demonstrate both numerically and experimentally in microwaves how we can easily build such a superdirective antenna. This strategy does not limit to the case of wires in microwaves and we give a very general recipe that can be applied to any wave field as long as one can find  subwavelength resonators.

Let us come back to the theoretical prediction of~\citet{schelkunoff_mathematical_nodate} to build a super-directive antenna. The idea is basically to impose the right distribution of currents to a series of sources. When packing such sources on a subwavelength scale you can manage to obtain a superdirective antenna~\cite{clemente_super_2014,clemente_design_2015}, but it requires a dedicated control of all the sources' currents as well as an optimization procedure since all the sources are strongly correlated. 
Our idea is therefore to use a single source but to add passive scatterers to achieve the super-directivity as it is done at larger scales for the case of the Yagi-Uda antennas~\cite{YagiUda1926}. Due to the subwavelength spacing between the wires we impose, we would intuitively anticipate that all scatterers would be in phase since the delay due to propagation is negligible, and therefore we should not be able to impose the right distribution of currents to fulfill the condition of~\citet{schelkunoff_mathematical_nodate}. However, metamaterials have demonstrated that it is possible to induce a change of sign between two neighbors whatever their separation distance~\cite{kaina_negative_2015} since the polariton branch can reach the edge of the first Brillouin zone~\cite{LemoultWRM2011a}. This tells us that the small dimension of our system is not a barrier for the use of passive scatterers for inducing directivity but at this stage we do not yet know how to impose the right distribution. The key relies on the resonant response of the scatterers. Indeed, the phase and amplitude of the current in a parasitic scatterer placed in the near-field of a point source strongly depends on the frequency. Typically, the phase response corresponds to an arctangent function where the phase shift occurs at the scatterer's eigenfrequency. and the multiple scattering problem can therefore exhibit some phase/amplitude differences between the distinct wires.

\begin{figure*}
\begin{center}
 \includegraphics[width=17cm]{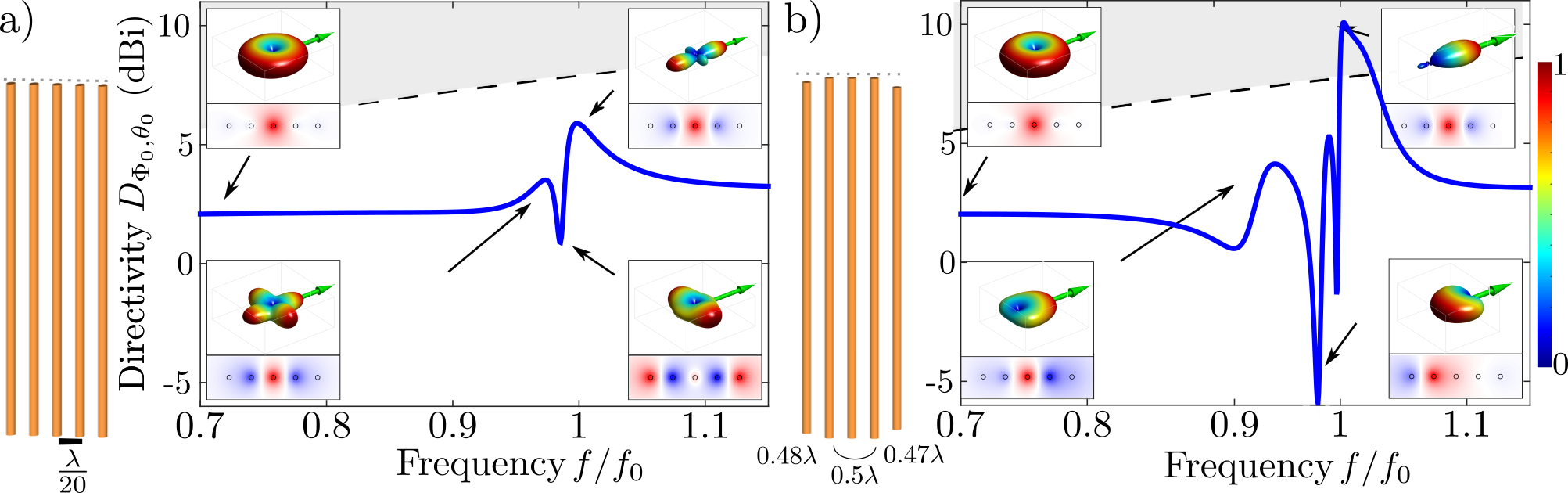}
 \caption{Analytical calculation of the directivity according to the frequency for \textbf{(a)} the periodic sample and \textbf{(b)} the optimized sample. In each panel, we represent the wire length distribution above the graph. Within the graph, the blue line corresponds to the frequency dependence of the directivity in the direction $(\phi_0,\theta_0)$. The black doted line is the calculated Harrington limit for this antenna size therefore the shaded area  corresponds to superdirectivity. The insets present the radiated farfield along with the analytical near field associated  at the frequencies indicated by the black arrows below. The optimized direction $(\phi_0,\theta_0)$ is pointed at by the red arrow. Calculation of the near field was made in a plane perpendicular to the wires, slightly above their upper end.} \label{figRP}
\end{center}
\end{figure*}

Fortunately, for the case of wires this multiple scattering problem can actually be solved quite easily by introducing the well documented impedance matrix~\cite{orfanidis}. The elements $Z_{mn}$ of this matrix are called the mutual impedance (or self-impedance if $m=n$), and they link the voltage induced in the wire $m$ when the wire $n$ is excited by a unitary current. Obviously the impedance matrix takes into account the frequency response of each wire through the current distribution along the wire. Note that this strategy does not limit to wires but can be generalized to any subwavelength resonators through the introduction of a $t$-matrix~\cite{Ad1998}. By inverting the impedance matrix and multiplying it by the excitation voltage vector, we retrieve the current that flows in each individual wire, {\it ie.} $I=Z^{-1}V$. Those currents then allow to compute the near field of the medium, and we eventually compute the radiated field by considering the superposition of the fields radiated by independent antennas fed by the previously calculated currents.       

As one can see in Fig.~\ref{fig_2D}, for a wire medium composed of 121 wires arranged on a square lattice so that it builds a cube of $\lambda/2$ a side, which is excited through the central wire, one can find analytically the dispersion relation from the multiple scattering inside of this cube. An easy way to understand the observed dispersive behaviour is to see it as an anti-crossing between the individual resonances of the wires and the freespace dispersion relation resulting in a polariton~\cite{hopfield1958,LemoultWRM2011a}, with an horizontal asymptote near $f_0$ exhibiting sub-wavelength varying modes. For frequency right above the resonance, no wave can propagate due to the existence of a so called hybridization bandgap~\cite{sigalas1993band}, which can equivalently be seen as a negative effective permittivity. This absence of propagation have been used to create cavities~\cite{kaina_ultra_2013} or waveguides~\cite{lemoult_wave_2013,kaina2017slow} with dimensions far below the freespace wavelength. The insets in the central part show near field calculations above the wires at different frequencies (red circles); one can see that the field distribution can be controlled on the scale of the wire separation rather than the scale of the freespace wavelength as the frequency rises. This confirms that metamaterials may be able to drag Schelkunoff's ideas into the realm of feasible experiments. However, as the symmetry of the system is yet to be broken, the emission of this small object is not yet superdirective.

In order to exhibit the effect of tuning the resonance of one scatterer, which in the case of wires solely consists in changing the length, as sketched in the right part of Fig.~\ref{fig_2D}, we tune one wire which allows one to use the hybridization bandgap to create a sub-wavelength cavity as in~\cite{kaina_ultra_2013}. Equally shortening a few of them allows one to engineer a waveguide as complicated as one wishes~\cite{lemoult_wave_2013,kaina2017slow}. Further control, as a gradient of eigenfrequencies, allows for even more interesting phenomenon, such as the rainbow trapping~\cite{,zhu_acoustic_2013,hu_rainbow_2013}. Those numerical maps were directly obtained by computing the multiple scattering problem inside the medium: this confirms that solely playing on the length of the passive wires can induce very different behaviour in the near-field. This will become our tuning parameter for controlling the local field distribution. 

For a simpler situation, let us now consider a linear array of five regularly spaced resonators with a separation distance of $d=\lambda/20$, where only the central wire is excited by a unitary voltage. The resulting currents that arises across the structure takes into account the whole multiple scattering, one is then hence able to compute the near-field by summing the contributions of all the secondary sources. Insets show the near-field at some key frequencies. The presence of those close resonant scatterers strongly interfere with the single source in the center, and one sees again a dispersive structuring of the near-field. Far from the resonance, the scattering cross-section is smaller and the scatterers barely interfere with the central source. Close to the resonance however, the near field shows a strong enhancement on the resonators, which means they are properly excited. They act as secondary sources and we can sum their coherent contributions to compute the farfield from the current distribution in the medium to study the radiation of this object. This farfield emission logically depends on the near-field structuring. We then focus on studying the directivity of the array. The term directivity is often used only for the main lobe direction, but here we define it relatively to a desired direction defined by the polar angles $(\theta_0,\phi_0)$, and it simply writes:
\begin{equation}
\label{dir}
D(\phi_0,\theta_0)=4 \pi\frac{P_r(\phi_0,\theta_0)}{\int_{0}^{2 \pi} \int_{0}^{\pi} P_r(\phi,\theta) \sin(\theta)\deriv \theta \deriv\phi}
\end{equation}

\noindent where $P_r$ is the measured radiated power in the direction $(\theta, \phi)$.

\begin{figure*}[bt]
\includegraphics[width=\textwidth]{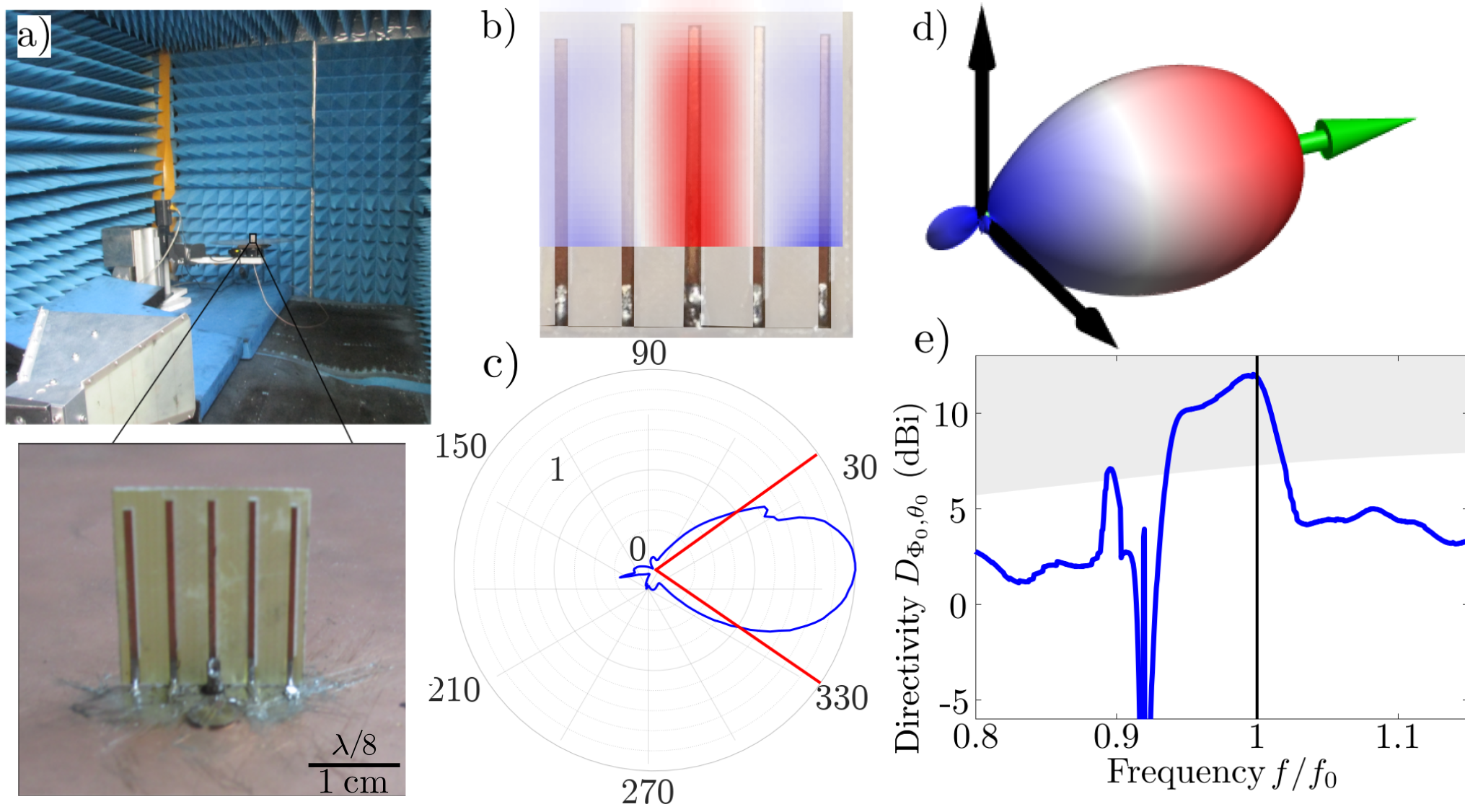}
\caption{(a) Experimental setup: inside the anechoic chamber the sample is placed on a 3D rotation stage consisting of two perpendicular rotational motors. A network analyzer is used to measure transmissions between samples and the receiving antenna. the bottom part shows a close up of the sample. (b) Near-field measurement of the current distribution on the meta-antenna. One sees the rapidly variation current distribution which corresponds to a Schelkunoff approach. (c) Polar representation of the radiated energy. The red lines gives the FWHM of about 57° (d) 3D representation of the radiation pattern. (e) Sample directivity according to frequency. Clear Superdirectivity of the sample is noticed over a bandwidth of more than 10\% of the central frequency.} \label{fig3}
\end{figure*}

In order to increase the directivity, a symmetry breaking is needed and our lever is the ability to tune the length distribution of the resonators. The question that arises is to know whether it is enough to control the current distribution in order to reach the super-directivity regime. Because of the self consistent problem at play, we developed a numerical strategy to find an optimized sample that maximizes the directivity in the desired direction. A discretized genetic algorithm is used to ensure rapid convergence and to avoid trapping effect of local optima. The parameter space is the length distribution confined to a 20\% relative difference to the central wire, discretized to the working precision of our sample construction method ($\lambda/160$).

An optimal length distribution is represented in Fig.~\ref{figRP}(b). Surprisingly, very small differences are enough to strongly alter the behaviour of the sample. Far from the resonance, we also find a dipolar-like radiation pattern. However, close to the resonance, the symmetry breaking yields asymmetric current distributions as one can see in the near fields. The corresponding radiated patterns highlight a variation of the main lobe direction, and a clearly superdirective peak, meaning a directivity above the Chu-Harrington limit (black dotted line), over a short bandwidth. One can also note that the directivity $D_{0,\frac{\pi}{2}}$ is negative at some points, meaning that the main lobe of emission at this frequency is pointing in another direction than the desired one as shown in the bottom right inset of Fig.~\ref{figRP}(b). As opposed to a simple Yagi-Uda design, here shorter wires are used on both ends of the array, showing the impact of multiple scattering. Indeed, this meta-antenna approach allows us to simplify Schelkunoff's idea, using a single source and the coupling with passive resonant scatterers to achieve the needed control for the current distribution rather than many sources where coupling will become a strong issue as one scales it down to sub-wavelength sizes. 
This control has then been demonstrated analytically, and the idea is now to experimentally verify this assertion by proceeding a microwave experiment of the concept.

The experimental samples are built as strips of copper (thickness 35~$\mu\textrm{m}$) over a thin layer of FR4 substrate ($\epsilon_r=4.2$), with a thickness of 0.4~mm. At the operating frequency of $3.74$~GHz the corresponding wavelength is of $8$~cm. We use a high precision drilling machine in order to accurately control separation and length of each strip. As mentionned, the discretization of lengths for the analytical samples was chosen to corresponds to this drilling precision. To avoid parasitic radiations from the cables in this experiment we preferred to shield the sources below a ground a plane. Therefore, we use half-long strips over a circular metallic ground plane. Those easy-to-build samples are accurately described by the previously discussed theory. A close-up of the realized sample is shown in Fig.~\ref{fig3}(a). The top part shows the experimental setup: measurements are performed in a home-made anechoic cavity using two perpendicular rotational motors under the sample to ensure $4\pi$-steradian measurements. The two antennas are connected to a network analyzer and the radiated field is measured in terms of the transmission between the two ports.

The experimental setup has two main effects over the resonance and radiated field. First, the FR4 substrate induces a frequency shift of the resonance towards lower frequencies. This shift has been correctly reproduced using transient simulations with the commercial software CST Microwaves Studio, and is solely due to the dielectric constant of the substrate. Second, the effect of a finite size ground plane on the radiated field of a monopole at its center is well documented and analytical formulas permit to predict the elevation angle dependence based on the ground plane radius~\cite{john_volakis_antenna_2007}. In this study, analytical results were obtained for coupled dipoles. To be consistent, we deconvolve the known azimuthal dependence of our samples to re-convolve the azimuthal dependence of analytical dipoles in post-processing. This actually lowers the measured directivity as the large ground plane increases antenna size and hence the accessible directivity. By doing so, we are actually measuring the emission of the corresponding dipoles while using shielded monopoles over a reflective plane. 

A near-field scan of the accessible part of a plane just 2 mm above the strips at the peak frequency is shown in Fig.~\ref{fig3}(b). One can notice its non-symmetric nature as well as its subwavelength varying nature ensuring a control of phase and amplitude of the secondary sources. A polar representation of the radiated energy is given in Fig.~\ref{fig3}(c). This corresponds to the polar measurement in the main-lobe direction of the monopoles. In Fig.~\ref{fig3}(d), one can see the complete radiation pattern for the corresponding dipoles, showing a strong directivity as high as 12 dB. As evidenced in the last panel, where the shaded region corresponds to superdirectivity, this experimental sample beats the Harington's limit over a relatively wide bandwidth of almost 10~\% of the central frequency. 

In conclusion, we have used an oversimplified wire medium to bring up to date Schelkunoff's theory of superdirectivity. Using the fact that  frequency detuning of some scatterers offer a lever to control the field distribution whatever the spacing between the wires, we have demonstrated both analytically and experimentally that it is possible to beat the Chu-Harington's limit, and we have therefore built a superdirective meta-antenna using only a single source with a well controlled environment that offers multiple scattering. The simplicity of the model can be extended to any scatterers or different wavelength, making it a crucial step for building small but directive antennas in any wave fields such as optics or even acoustics.

%

%



\end{document}